\documentclass[twocolumn,superscriptaddress,longbibliography,aps,preprintnumbers]{revtex4-2}

\usepackage{graphicx,color}
\usepackage{xcolor}
\usepackage{bm,bbm,amsmath,amssymb,braket}
\usepackage{epstopdf}
\usepackage{relsize}
\usepackage{epstopdf}
\usepackage{comment}
\usepackage{soul}
\usepackage{float}
\usepackage{lipsum}
\usepackage[urlcolor=blue,colorlinks=true,citecolor=blue,
	linkcolor=blue,pdfstartview={FitH},bookmarks=false]{hyperref}
\usepackage{MnSymbol}
\usepackage{stmaryrd}
\usepackage{mathtools}

\graphicspath{{Figures/}{./Figures/}{.}{./paper_figures/}}
\setstcolor{blue}

\definecolor{purple}{rgb}{0.5, 0., 0.8}

\definecolor{persianblue}{rgb}{0.11, 0.22, 0.73}

\definecolor{darkred}{RGB}{182, 24, 24}

\begin{document}
\title{General many-body entanglement swapping protocol:  opportunities for distributed quantum computing}
\author{Santeri Huhtanen}
\affiliation{Computational Physics Laboratory, Physics Unit, Faculty of Engineering and Natural Sciences, Tampere University, P.O. Box 692, FI-33014 Tampere, Finland}
\affiliation{Helsinki Institute of Physics P.O. Box 64, FI-00014, Finland}
\author{Yousef Mafi}
\affiliation{Computational Physics Laboratory, Physics Unit, Faculty of Engineering and Natural Sciences, Tampere University, P.O. Box 692, FI-33014 Tampere, Finland}
\affiliation{Helsinki Institute of Physics P.O. Box 64, FI-00014, Finland}
\author{Ali G.  Moghaddam}\email{Email: ali.moghaddam@aalto.fi}
\affiliation{Department of Applied Physics, Aalto University, 02150 Espoo, Finland}
\affiliation{Computational Physics Laboratory, Physics Unit, Faculty of Engineering and Natural Sciences, Tampere University, P.O. Box 692, FI-33014 Tampere, Finland}
\affiliation{Helsinki Institute of Physics P.O. Box 64, FI-00014, Finland}
\author{Teemu Ojanen}\email{Email: teemu.ojanen@tuni.fi}
\affiliation{Computational Physics Laboratory, Physics Unit, Faculty of Engineering and Natural Sciences, Tampere University, P.O. Box 692, FI-33014 Tampere, Finland}
\affiliation{Helsinki Institute of Physics P.O. Box 64, FI-00014, Finland}

\date{\today}

\begin{abstract}
Sharing entangled pairs between non-signaling parties via entanglement swapping constitutes a striking demonstration of the nonlocality of quantum mechanics and a crucial building block for future quantum technologies. In this work, we generalize pair-swapping methods by introducing a many-body entanglement swapping protocol, which allows two non-signaling parties to share general many-body states along an arbitrary partitioning. The shared many-body state retains exactly the same Schmidt vectors as the target state and exhibits typically high fidelity, which approaches unity as the variance of the Schmidt coefficients vanishes. 
Moreover, we demonstrate how the three-party protocol can be generalized to many-body swapping networks, enabling a general many-body state sharing with unit fidelity via arbitrary number of intermediate nodes.
This is achieved by replacing all but one of the unitary operations with those corresponding to the same Schmidt states but with a flattened spectrum, which also completely eliminates the need for postselection. We provide a proof of concept of the three-party protocol on real quantum hardware and discuss how it enables new functionalities, such as  fault-tolerant entanglement swapping and new strategies for distributed quantum computing.                

\end{abstract}

\maketitle

\section{Introduction}

The ability to share quantum states nonlocally—across distances ranging from microns to hundreds of kilometers—is one of the most intriguing features of quantum mechanics. At the same time, it plays a central role in enabling practical tasks in quantum information processing, particularly in distributed systems and quantum networks \cite{Cirac1997Distribution,Kimble2008}. One powerful protocol for achieving long-range quantum information transfer is \emph{entanglement swapping}, which allows two distant, non-signaling parties to share an entangled state through the use of intermediate entangled pairs \cite{zukowski1993event,Zeilinger2001}. This counterintuitive phenomenon highlights the fundamentally nonlocal character of quantum mechanics and represents a clear departure from classical intuition \cite{Yurke1992,brukner2005complementarity,Gisin2005swapping}. Beyond its conceptual significance, entanglement swapping has become a key tool in the development of quantum repeaters, quantum communication architectures, and foundational tests of quantum mechanics \cite{Horodecki2009,Quantum_repeaters2023,bose1998swapping,Pan2009MultiparticleSwapping,bej2020swapping,Jing2022Swapping}.

In addition to long-distance applications, the ability to share quantum states over short distances is equally critical for quantum computing—especially as we look beyond the Noisy Intermediate-Scale Quantum (NISQ) era. This current era has been constrained by limited qubit counts, high error rates, and sparse hardware connectivity \cite{preskill2018NISQ,divincenzo2000physical,corcoles2019challenges,georgescu2020divincenzo}. Looking ahead, the envisioned “megaquop” era anticipates quantum processors capable of reliably performing on the order of one million operations \cite{preskill2025megaquop}. In this context, the need for robust and scalable methods to generate and control complex entanglement across distant qubits becomes even more pressing.
Such capabilities support not only nonlocal multi-qubit gates—essential for advanced algorithms and quantum error correction \cite{bravyi2024ldpc}—but also play a foundational role in quantum simulation \cite{fauseweh2024quantum}, distributed computing models, and the development of large-scale quantum supercomputers \cite{mohseni2025supercomputer}.

\begin{figure}[t]
    \centering
    \includegraphics[width=.99\columnwidth]{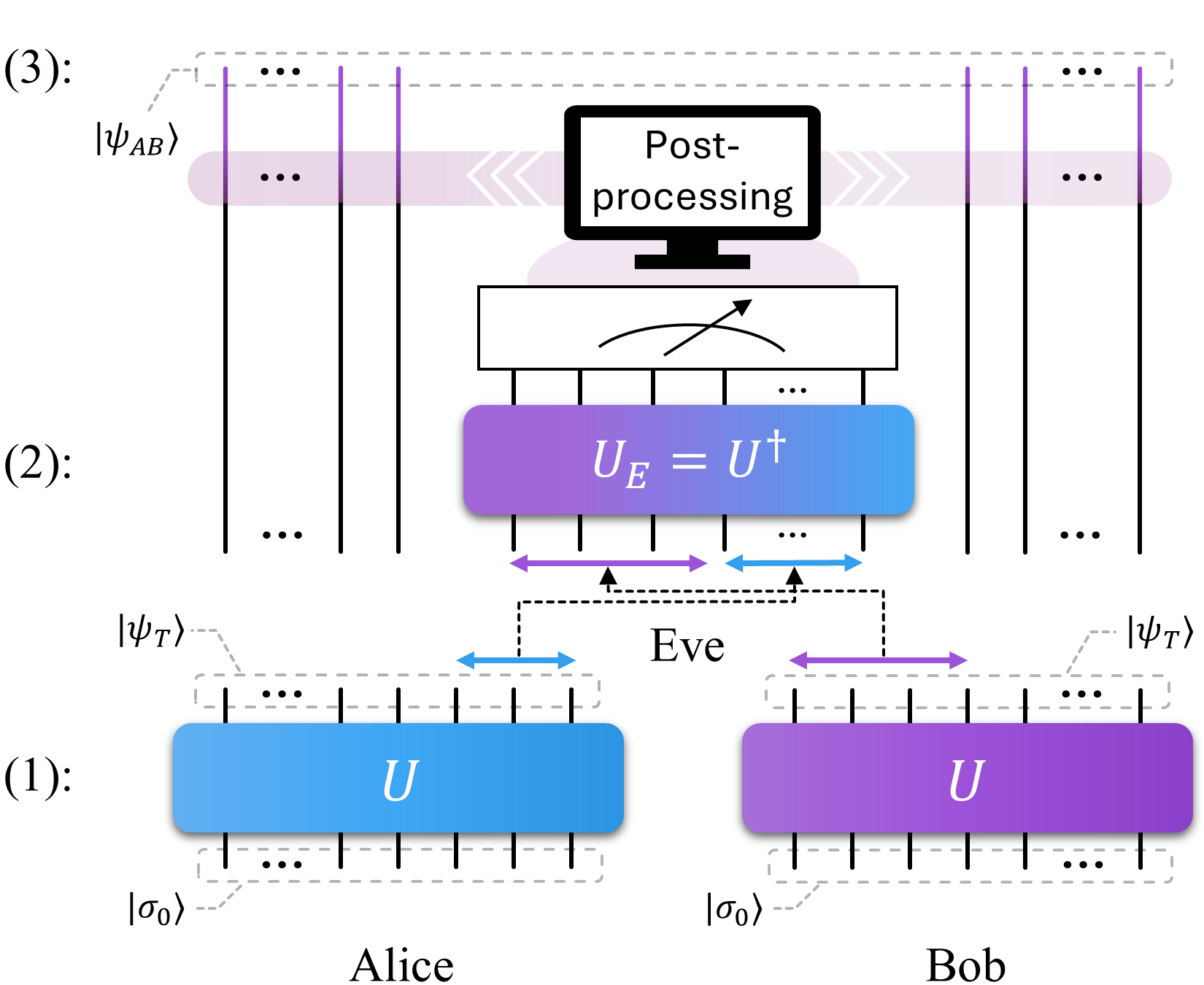}  
        \caption{ Many-body entanglement swapping protocol. In step 1, Alice and Bob create locally a target state $|\psi_T\rangle= U|\sigma_0\rangle$ which determines the entanglement structure of the shared state. They share the state along partition in which Alice holds $n_A$ and Bob holds $n_B$ qubits respectively, and send the remaining qubits to Eve. In step 2, Eve applies unitary $U_E$ to her qubits and then measures them. If she obtains result $|\sigma_0\rangle$, she informs Alice and Bob to keep their qubits, otherwise discard them. In step 3, after postselection, Alice and Bob share entangled state $|\psi_{AB}\rangle$ which share exactly the same Schmidt eigenstates as $|\psi_{T}\rangle$ and simply related Schmidt coefficients $\lambda_i^{AB}$.   }
    \label{fig1}
\end{figure}

In this work, we introduce a many-body entanglement‐swapping protocol which generalizes pair-swapping methods and enables flexible, high‐fidelity sharing of complex multi-qubit states $|\psi_{AB}\rangle$. As illustrated in Fig.~\ref{fig1}, the protocol proceeds in three stages. First, Alice and Bob each generate the target state $|\psi_T\rangle$ with a unitary $U$, retain the qubits corresponding to their respective partitions, and transmit the remaining qubits to an intermediary party, Eve. In the second stage, Eve applies the inverse unitary $U^\dagger$ to the received qubits, performs a projective measurement, and broadcasts the outcome to Alice and Bob. When Eve’s measurement yields a desired result, Alice and Bob’s joint state collapses to $|\psi_{AB}\rangle$, which shares exactly the same Schmidt vectors as $|\psi_T\rangle$, and whose Schmidt coefficients are simply related to those of the target state. We show that the fidelity of the shared states is typically high—even for complex many-body states—and becomes unity when the Schmidt spectrum is uniform.  We then illustrate how the three-party protocol can be generalized to many-body swapping networks, where Alice and Bob share a state via arbitrary number of intermediate nodes. The fundamental cost of the swapping, characterized by the postselection probability in intermediate nodes, is determined by R\'enyi entanglement entropies of the partitioning. Moreover, we demonstrate that postselection can be entirely eliminated, allowing the states to be shared with unit fidelity, by replacing the unitary operator at the final intermediate node of the network with another unitary that preserves the same Schmidt basis while possessing a flattened (i.e., uniform) spectrum. We demonstrate the proof of principle of the three-party protocol on real quantum hardware and explain how it allows fault-tolerant quantum state sharing between non-signaling parties.

The many-body swapping protocol, enabling a reliable sharing of complex entangled states, opens new avenues for scalable quantum networks and distributed quantum information processing. Particularly, while recent advances in dynamic quantum circuits, featuring mid-circuit measurements and real-time feed-forward, have enabled the teleportation of two-qubit gates across large superconducting arrays \cite{Minev2024,Seif2024dynamic, carrera2024combining}, these approaches are mostly effective for low-complexity states and become increasingly inefficient for more intricate entanglement structures. In contrast, our many-body swapping protocol addresses these limitations by offering a scalable and hardware-efficient strategy for entanglement distribution, which is essential for the megaquop era and large-scale quantum technologies.

\section{Many-body entanglement swapping protocol} 

\subsection{Many-body swapping via single intermediary} \label{subsec:single}

In this section we outline our many-body entanglement swapping protocol. We search for a method that would, on one hand, allow two non-signaling parties Alice and Bob to share a target many-body state $|\psi_T\rangle$, which they can generate locally, along an arbitrary partitioning. On the other hand, not counting the usual practical limitations of carrying out unitary transformations and measurements, the protocol should be explicit and straightforwardly applicable. While the ability to share a state precisely and straightforwardly are competing demands, below we see that it is possible to reconcile with the requirement to an impressive degree. We identify a large class of systems where the sharing can be done exactly or with high accuracy, and even when that is not the case, the entanglement structure of the shared state is straightforwardly inherited from the target state.   

The general protocol is agnostic to the precise nature of the elementary degrees of freedom (qubits, qudits etc), but for the sake of concreteness, we consider a system of qubits. Assume that two parties, Alice and Bob, can locally prepare a target multi-qubit state 
\begin{align}\label{eq:target}
|\psi_T\rangle=U|\sigma_0\rangle,    
\end{align}
where $U$ is unitary acting on $n$ qubits and $|\sigma_0\rangle$ denotes a product state in the computational basis. Now, Alice and Bob would like to share $|\psi_T\rangle$ along an arbitrary partitioning in such a way that Alice is in possession of a set of qubits $\{N_A\}$  with $n_A$ elements, spanning the Hilbert space $\mathcal{H}_A$. Bob, on the other hand, should be in the possession of the remaining set $\{N_B\}$ with $n_B=n-n_A$ elements, spanning the Hilbert space $\mathcal{H}_B$. Furthermore, the protocol should be carried out without signaling between Alice and Bob, but with a help of a third party, Eve, as illustrated in Fig.~\ref{fig1}.

In the first step, both parties prepare locally the target state. Thus, the initial state becomes 
\begin{equation}
|\Psi_1\rangle=|\psi_T\rangle\otimes|\psi_T\rangle, \nonumber
\end{equation}
where $|\psi_T\rangle\in \mathcal{H}_A\otimes \mathcal{H}_B$.  
Considering a bipartitioning into subsystems $\{N_A\}$ and $\{N_B\}$, the Schmidt decomposition of the target state can be written as 
\begin{align}\label{eq:schmidt1}
    |\psi_T\rangle=\sum_{i} \sqrt{\lambda_i}|\lambda_i^A\rangle\otimes |\lambda^B_i\rangle,
\end{align}
where $\lambda_i>0$ are the Schmidt coefficients and $|\lambda_i^A\rangle\in \mathcal{H}_A$, $|\lambda_i^B\rangle\in\mathcal{H}_B$ are the orthonormal Schmidt vectors corresponding to the partitioning. Using this, we can express the initial state of the two parties as   
\begin{equation}
|\Psi_{1}\rangle=\sum_{i,j} \sqrt{\lambda_i\lambda_j}|\lambda^A_i\rangle\otimes |\lambda^B_i\rangle\otimes|\lambda^A_j\rangle\otimes|\lambda^B_j\rangle. \label{eq:step1-swap-basic}
\end{equation}

In the second step, Alice sends her set of qubits $\{N_B\}$ and Bob his set $\{N_A\}$ to Eve, who will then apply a unitary $U_E$ on her qubits, leading to state $|\Psi_{2}\rangle=U_E|\Psi_{1}\rangle$. This can be expressed as 
\begin{equation}\label{eq:stage2}
|\Psi_{2}\rangle=\sum_{i,j} \sqrt{\lambda_i\lambda_j}|\lambda_i^A\rangle\otimes U_E \left(|\lambda^B_i\rangle\otimes|\lambda^A_j\rangle\right)\otimes|\lambda_j^B\rangle.\nonumber
\end{equation}
Then, Eve measures all her qubits in the computational basis $\{|\sigma_i\rangle\}$. Assuming the measurement outcome is $|\sigma_E\rangle$, the final state becomes 

 \begin{align}
|\Psi_{3}\rangle &=\frac{1}{\sqrt{p_E}} \left( \mathbbm{1} \otimes |\sigma_E\rangle \langle\sigma_E|\otimes \mathbbm{1}\right) \ket{\Psi_{2}} \nonumber\\
&= \frac{1}{\sqrt{p_E}} \sum_{i,j} \sqrt{\lambda_i\lambda_j}\, r_{ij}^E \:
|\lambda^A_i\rangle\otimes |\sigma_E\rangle\otimes|\lambda^B_j\rangle,\nonumber
\end{align}
where the coefficients $r_{ij}^E$ are 
\begin{equation}\label{eq:rij}
r_{ij}^E\equiv\langle \sigma_E|U_E\left(|\lambda^B_i\rangle\otimes|\lambda^A_j\rangle\right),   \nonumber
\end{equation}
and the probability of outcome $|\sigma_E\rangle$ is given by
\begin{equation}\label{eq:prob}
p_E=\sum_{i,j}\lambda_i\lambda_j|r_{ij}^E|^2.
\end{equation}
As the final state has a product form with respect to Eve, her state can be trivially factored out. Thus, we can focus on the joint $n$ qubit state for which Alice holds $n_A$ and Bob $n_B$ qubits:
\begin{align}\nonumber
|\psi_{AB}\rangle = \frac{1}{\sqrt{p_E}} \sum_{i,j} \sqrt{\lambda_i\lambda_j}\, r_{ij}^E \:
|\lambda^A_i\rangle\otimes |\lambda^B_j\rangle.
\end{align}
The joint state of Alice and Bob depends on Eve's measurement outcome $|\sigma_E\rangle$ in the last step, which is nondeterministic. This fact, in general, necessitates postselection, which can be avoided in special cases. The overlap of the target state and the state resulting from the many-body swapping becomes 
\begin{equation}\label{eq:overlap1}
F=\langle\psi_T |\psi_{AB}\rangle= \frac{ 1}{\sqrt{p_E}}\sum_{k} \lambda_k^{3/2}r_{kk}^E ,
\end{equation}
providing an important figure of merit $|F|$ called fidelity, which satisfies $0\leq |F|\leq 1$. Ideally, one would like $|F|$ to be as close to unity as possible while satisfying the constraint $\sum_{ij}|r_{ij}^E|^2\leq 1$. Optimal solutions clearly satisfy $r_{ij}^E=r_{ii} \delta_{ij}$ and all $r_{ii}$ should have the same complex phase. These two conditions now suggest an efficient strategy for Eve to choose her unitary $U_E$  and the postselected state. In above, we assumed that $U_E$ is acting in the Hilbert space $\mathcal{H}_B\otimes\mathcal{H}_A$, but it is now convenient to work in the reordered basis $\mathcal{H}_A\otimes\mathcal{H}_B$, where the corresponding unitary is denoted as $U_E'$. The two representations are related so that $U_E(|\lambda_i^B\rangle\otimes |\lambda^A_i\rangle)$ in the basis $\mathcal{H}_B\otimes\mathcal{H}_A$  corresponds to the vector $U'_E(|\lambda_i^A\rangle\otimes |\lambda^B_i\rangle)$ in the basis $\mathcal{H}_A\otimes\mathcal{H}_B$
for all vectors $|\lambda^A_i\rangle\in \mathcal{H}_A$ and $|\lambda^B_i\rangle\in \mathcal{H}_B$. Combining Eqs.~\eqref{eq:target},~\eqref{eq:schmidt1}, we see that 
\begin{align}\label{eq:schmidt2} 
   U |\sigma_0\rangle=\sum_{i} \sqrt{\lambda_i}|\lambda_i^A\rangle\otimes |\lambda^B_i\rangle.\nonumber
\end{align}
Thus, if Eve chooses $U_E'=U^\dagger$ and postselects to retain only the measurement outcomes for which $|\sigma_E\rangle=|\sigma_0\rangle$, we see that $r_{ij}^E=\sqrt{\lambda_i}\delta_{ij}$ and $r_{ii}>0$. The fidelity then becomes 
\begin{equation}\label{eq:overlap2}
F= \frac{\sum_{k}\lambda_k^2 }{\sqrt{\sum_{m}\lambda_m^3}}=e^{S_3(\{\lambda_i\})-S_2(\{\lambda_i\})},
\end{equation}
where we have employed R\'enyi entropies 
\begin{equation}\label{eq:renyi}
S_n(\{\lambda_i\})=\frac{1}{1-n}\ln \sum_{i}\lambda_i^n.\nonumber
\end{equation}
 After the swapping protocol, the shared state itself becomes
 \begin{equation}\label{eq:final2}
|\psi_{AB}\rangle=\sum_{i} \sqrt{\lambda^{AB}_i}|\lambda^A_i\rangle\otimes|\lambda^B_i\rangle,
\end{equation}
which has the exact same Schmidt vectors than the target state. The Schmidt coefficients of the shared state are simply given in terms of the target state coefficients as 
 \begin{equation}\label{eq:schmidt2}
\lambda^{AB}_i=\frac{\lambda_i^3}{\sum_{m}\lambda_m^3}.
\end{equation}
The probability \eqref{eq:prob} of obtaining the desired outcome state $|\sigma_0\rangle$ 
in Eve's measurement in turn becomes 
 \begin{equation}\label{eq:prob1}
p_0=\sum_{i}\lambda_i^3=e^{-2S_3(\{\lambda_i\}) }.
\end{equation}
This probability is also the success rate for the postselection and characterizes the mean number of trials $\sim 1/p_0$ to share a single copy of $|\psi_{AB}\rangle$ between Alice and Bob. 

A few observations are in order. The overlap~\eqref{eq:overlap2} between the shared and target states reaches unity when 
$S_2(\{\lambda_i\}) = S_3(\{\lambda_i\})$. 
This condition is satisfied when the Schmidt spectrum is uniform, i.e., $\lambda_i = \lambda_j$ for all $i, j$. 
This situation covers important classes of quantum states namely \emph{maximally entangled states} and also \emph{stabilizer states}, which include GHZ-type and cluster states, which play a central role in quantum information processing and quantum computation. 
In particular, stabilizer states possess Schmidt coefficients of the form $\lambda_i = 2^{-r}$, with $r \in \mathbb{N}$, and their Schmidt vectors are themselves stabilizer states~\cite{fattal2004}. 
Moreover, the set of unit-fidelity states also encompasses a broad class of \emph{non-stabilizer states} characterized by uniform Schmidt coefficients $\lambda_i = 1/q$, where $q \in \mathbb{N}$, and arbitrary Schmidt vectors.

Importantly, the overlap often remains very high even for non-uniform spectrum. To see this, it is useful to parameterize the Schmidt coefficients as $\lambda_i=d_S^{-1}+\epsilon_i$, where $d_S$ is the Schmidt rank (the number of nonzero Schmidt coefficients) and $\epsilon_i$ is the deviation from the mean value. The fidelity \eqref{eq:overlap2} then becomes 
\begin{equation}\nonumber
F= \frac{1+d_S^2\overline{\epsilon^2} }{\sqrt{1+3d^2_S\overline{\epsilon^2}+d^3_S\overline{\epsilon^3}}}\xrightarrow[\overline{\epsilon^2} \to 0]{} 1-\frac{d_S^2\overline{\epsilon^2}}{2}+\mathcal{O}(d_S^3\overline{\epsilon^3}),
\end{equation}
where $\overline{\epsilon^n}=d^{-1}_S\sum_i\epsilon_i^n$ for $n=2,3$ determine the variance and the skewness of the Schmidt coefficients. Thus, the overlap remains nearly perfect whenever the non-uniformity satisfies $d_S^2\overline{\epsilon^2}\ll1$. Even for significant non-uniformity $d_S^2\overline{\epsilon^2},d_S^3\overline{\epsilon^3}\sim1$, the fidelity remains remarkably high $F\sim 2/\sqrt{5}\sim 0.9$. Below we illustrate how complex multi-qubit random states can be shared with comparable fidelity. In addition, irrespective of the fidelity, the entanglement structure of the shared state is always inherited from the target state, with exactly coinciding Schmidt vectors and the Schmidt coefficients that are related to those of the target state by Eq.~\eqref{eq:schmidt2}. Therefore, the protocol enables an explicit method of sharing a known many-body state with a close resemblance to a general target state. 
    
Finally, the fundamental cost of the many-body swapping is determined by the success probability \eqref{eq:prob1} of postselection. The lower the probability, the more repetitions are required for sharing a single copy of $|\psi_{AB}\rangle$. Importantly, the cost  \emph{does not} depend explicitly on the number of qubits in the shared state, but only on the entanglement entropy $S_3$ of the partitioning. Thus, the fundamental protocol complexity does not directly depend on the system size, but only on the shared entanglement. For example, many interesting one-dimensional condensed-matter phases and their quantum circuit representations exhibit an area-law entanglement entropy scaling \cite{RevModPhys.82.277, RevModPhys.93.045003}. This means that the entanglement entropy of partitioning does not depend on the total system size, so the cost of sharing them would also be size independent. 
The same conclusion also holds for 1d states generated by \emph{sequential quantum circuits} capable of generating area-law entangled states with complex internal structure, including GHZ states and various topologically ordered states \cite{Cirac2022,Zhao2024Sequential}.
In some simple cases, as discussed below, it is possible to avoid postselection simply by applying local unitaries to correct the shared state, as in the pair swapping.

\subsection{Many-body swapping network}
\label{subsec:multiple}

\begin{figure}[t]
    \centering \includegraphics[width=.99\columnwidth]{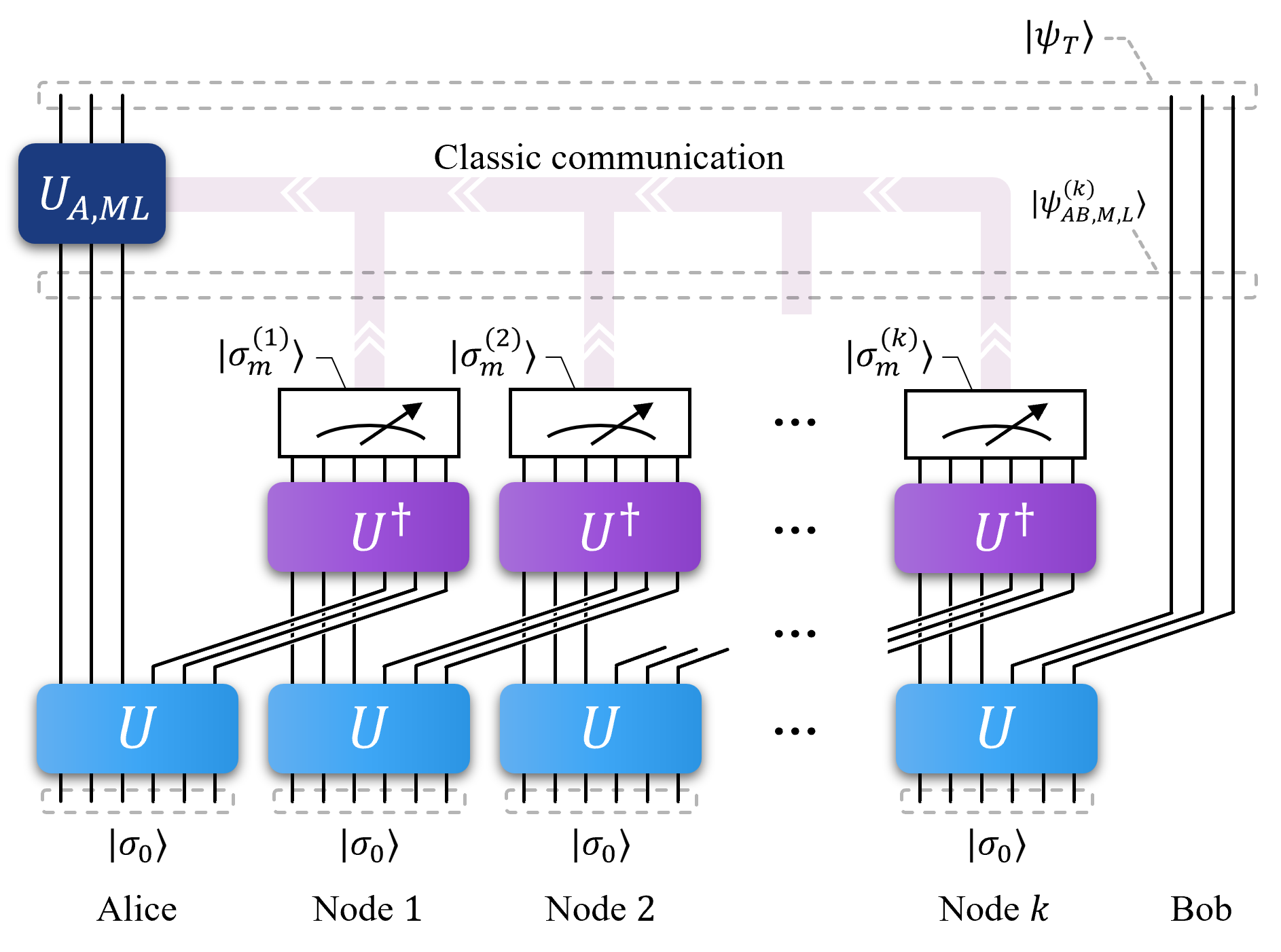}  
    \caption{Sharing many-body states via many-body swapping in a network with $k$ intermediate nodes. The protocol illustrated here corresponds to the postselection-free implementation discussed in Subsec.~\ref{subsec:postselection}. In the original version of the protocol (see Subsec.~\ref{subsec:multiple}), postselection was required based on the measurement outcomes at all intermediate nodes. However, as explained in Subsec.~\ref{subsec:postselection}, when the Schmidt spectrum is uniform, postselection can be entirely avoided. In this case, each node simply records its measurement outcome and communicates it to Alice, who then applies an appropriate local unitary operation $U_{A,ML}$ on her remaining qubits, resulting in a shared state between Alice and Bob with unit fidelity. Finally, as described in Subsec.~\ref{subsec:general_nonpostselection}, a similar approach can be applied to states with a non-uniform Schmidt spectrum. }
    \label{fig:network}
\end{figure}
The above discussed many-body state sharing via single node, Eve, can be generalized to state sharing via arbitrary number of intermediate nodes. This is schematically illustrated in Fig.~\ref{fig:network}. The goal of the state sharing is the same as above, Alice and Bob would like to share the target state $|\psi_T\rangle$ along an arbitrary partition so that Alice has a set of qubits $\{N_A\}$  with $n_A$ elements while Bob has the remaining set $\{N_B\}$ with $n_B=n-n_A$ qubits. In this scheme, Alice and each intermediate nodes locally generate the shared state, keep the $n_A$ qubits that belong to Alice's partition, and send the $n_B$ qubits in the Bob's partition to the next node in the network. The state sharing through multiple nodes is conveniently analyzed recursively by analyzing how the shared state $|\psi_{AB}^{(k-1)}\rangle$ of a network with $k-1$ nodes evolves into the state shared via $k$ nodes. The result for a single node $k=1$ is mathematically equivalent to the state sharing protocol studied in the previous subsection, so
\begin{equation}\nonumber
 |\psi_{AB}^{(1)}\rangle=|\psi_{AB}\rangle,   
\end{equation}
where $|\psi_{AB}\rangle$ is given by Eq.~\eqref{eq:final2}. The only difference of a single node protocol and the state sharing via Eve discussed in Sec.~\ref{subsec:single} is that previously we considered a situation where Alice and Bob both generated $|\psi_T\rangle$ and sent qubits to Eve for processing. Here the node generates the state and sends $n_B$ qubits to Bob before carrying out the operation with $U^\dagger$ followed by the measurement. This difference does not affect the shared state of Alice and Bob. The intermediate node performs the same postselection as Eve, described by the success probability \eqref{eq:prob1}.

Let's now add a second intermediate node and work out the shared state of Alice and Bob $|\psi_{AB}^{(2)}\rangle$  in this case. The key observation is that the state of the system prior to the second swapping can be expressed as the product of
\begin{enumerate}
    \item the shared state of Alice and node $1$, and 
    \item the target state generated at node $2$ by applying the third $U$ operation,
\end{enumerate}
and, thus, it can be written as
\begin{align}
|\Psi\rangle = |\psi_{AB}^{(1)}\rangle \otimes |\psi_T\rangle.
\end{align}
Next, $n_B$ qubits from the second node are transmitted to Bob, and the second unitary operation $U^\dagger$ is applied to the combined system consisting of the remaining $ n_A $ qubits and the $ n_B$ qubits received from the first node.
The only difference to the single-node calculation presented in the previous subsection is that $|\psi_{AB}^{(1)}\rangle$ now has Schmidt coefficients \eqref{eq:schmidt2} instead of $\lambda_i$. Due to this difference, the shared state 
between Alice and Bob after second postselection becomes
\begin{equation}
 |\psi_{AB}^{(2)}\rangle=\sum_{i} \sqrt{\lambda^{(2)}_i}|\lambda^A_i\rangle\otimes|\lambda^B_i\rangle,   
\end{equation}
where $\lambda^{(2)}_i=\lambda_i^5/\sum_j\lambda_j^5$. These results are now straightforward to generalize to $k$ nodes. The initial state before the last swapping process 
between the nodes $k$ and $k-1$ is just a product of the shared state of Alice and $(m-1)$\textsuperscript{th} node and the state generated in the $m$\textsuperscript{th} node   
\begin{align}\label{eq:mthswap}
|\Psi\rangle=|\psi_{AB}^{(k-1)}\rangle\otimes|\psi_T\rangle.    
\end{align}
Then, $n_B$ qubits from the last node are send to Bob, and $k$\textsuperscript{th} swapping operator $U^\dagger$ is applied to the remaining qubits from the nodes $k$ and $k-1$, followed by 
measuring them. After postselection in the last measurement, the shared by Alice and Bob becomes 
\begin{equation}\label{eq:final_m}
 |\psi_{AB}^{(k)}\rangle=\sum_{i} \sqrt{\lambda^{(k)}_i}|\lambda^A_i\rangle\otimes|\lambda^B_i\rangle,   
\end{equation}
where the Schmidt coefficients are directly given by those of the target state as
 \begin{equation}\label{eq:schmidt3}
\lambda^{(k)}_i=\frac{\lambda_i^{2k+1}}{\sum_{j}\lambda_j^{2k+1}}.
\end{equation}
With this result, one can immediately generalize the results for the single-node fidelity \eqref{eq:overlap2} and the postselection probability \eqref{eq:prob1} to corresponding quantities in the many-body swapping through $k$ intermediate nodes. Expressed in terms of the R\'enyi entropies, the multinode fidelity becomes

\begin{equation}\label{eq:overlap3}
F^{(k)}= \frac{\sum_{i}\lambda_i^{k+1} }{\sqrt{\sum_{i}\lambda_i^{2k+1}}}=e^{k\left[ S_{2k+1}(\{\lambda_i\})-S_{k+1}(\{\lambda_i\})\right]},
\end{equation}
while the postselection success probability at the $k$th node takes the form
\begin{align}\label{eq:prob2}
  p_0^{(k)}=\frac{\sum_i\lambda_i^{2k+1}}{\sum_{i}\lambda_i^{2k-1}}=e^{2k\,S_{2k+1}-2(k-1)\,S_{2k-1}}.
\end{align}
Thus, the general $k$ node results reduce to those of the previous subsection when $k=1$. The swapping over the network proceeds to the next node if the measurement in the previous node provided the desired result $|\sigma_0\rangle$, otherwise the node where the failure happened terminates the protocol and informs Alice to discard the state.  If Alice does not receive a notification of a failure, Alice and Bob know they the share state \eqref{eq:final_m}.

It should be emphasized that, as in the single node case, all many-body states with constant Schmidt coefficients satisfy $S_{2k+1}(\{\lambda_i\})=S_{k+1}(\{\lambda_i\})$ and \emph{can be shared exactly with unit fidelity}. As discussed in the previous subsection, these states include key classes of states for quantum computing. As discussed in Sec.~\ref{subsec:DistributedQC} below, this enables much-needed and highly sought-after strategies to implement distributed quantum computing. Remarkably, as proved in the next subsection, these states can be swapped without any postselection cost.

\subsection{Avoiding postselection for states with uniform Schmidt coefficients} \label{subsec:postselection}

The results in the preceding subsections demonstrate how many-body quantum states can be shared by many-body swapping protocol. The postselection, in general, poses a bottleneck for setting up a scalable quantum network or sharing highly entangled states. Remarkably, this bottleneck can be completely avoided. In this section we show how this can be carried out for states with uniform Schmidt coefficients and, in the following section, how this method can be applied to general states. This would enable, for example, efficient sharing of multi-qubit stabilizer states between distinct modular quantum processors units as discussed in Sec.~\ref{subsec:DistributedQC}.

The unit fidelity states are in general of the form 
\begin{equation}\label{eq:final_m}
 |\psi_{T}\rangle=\sum_{i=0}^{d_S} \frac{1}{\sqrt{d_S}}|\lambda^A_i\rangle\otimes|\lambda^B_i\rangle,\end{equation}
where the Schmidt basis states $\{|\lambda^A_i\rangle\}$, $\{|\lambda^B_i\rangle\}$ form two orthonormal sets. The R\'enyi entropies of \eqref{eq:final_m} are given by $S_n=\log d_S$, so the postselection probability in state sharing over a network of $k$ nodes becomes exponentially suppressed as $\prod_{q=1}^{k} p_0^{(q)}=d_S^{-2k}$. Thus, it is important to develop a method to avoid postselection altogether. 

In formulating the many-body swapping protocol, we assumed that the target state $|\psi_T\rangle$ is generated by unitary $U$ from a computational basis state $|\sigma_0\rangle$ according to Eq.~\eqref{eq:target}. But, this does not fix $U$ uniquely, leaving a lot of room for choosing a specific form of $U$. This vast freedom can be exploited to avoid postselection by considering $U$ such that it maps different computational basis states to mutually orthogonal \emph{target-like} states, which can be transformed into the exact target state by local operations by Alice and Bob. For a reason which will become clear shortly, we fix the remaining freedom in $U$ so that it maps $d_S^2$ different computational basis states into such target-like states. For a given target state   
$|\psi_{T}\rangle$ with the Schmidt form in Eq. \eqref{eq:final_m}, shifting the relative labels of two sets of Schmidt vectors $|\lambda_{i}^A\rangle$  and $|\lambda_{i}^B\rangle$ as the following, would give different target-like states
\begin{align}\label{eq:basis_shift}
 |\psi_{m,0}\rangle=\sum_{j=0}^{d_S-1} \frac{1}{\sqrt{d_S}}|\lambda^A_j\rangle\otimes|\lambda^B_{j+m}\rangle,
 \end{align}
where the label $j+m$ should be considered cyclic i.e. $\mod d_S$. Then, by considering Fourier modulated linear combinations
\begin{align}\label{eq:basis_shift_comp}
 |\psi_{m,l}\rangle=\sum_{j=0}^{d_S-1} \frac{e^{2\pi i j l/d_S}}{\sqrt{d_S}}|\lambda^A_j\rangle\otimes|\lambda^B_{j+m}\rangle,
 \end{align}
where $m,l \in \{0,1,\ldots d_S-1\}$, 
we get $d_S^2$ the needed orthonormal target-like states $\langle \psi_{m',l'}  |\psi_{m,l}\rangle=\delta_{m'm}\delta_{l'l}$. Now we require that the unitary operator $U$ employed in the swapping process generate these orthonormal states $|\sigma_{m,l}\rangle$ from a set of computational basis state $|\sigma_{m,l}\rangle$ as $U |\sigma_{m,l}\rangle = |\psi_{m,l}\rangle$. Thus, the generating unitary is simply chosen so that it transforms $d^2_S$ orthonormal basis states to the set $|\sigma_{m,l}\rangle$, which can be always arranged. The crucial point is that due to this construction, after each measurement, we always obtain an outcome which is of the form $|\sigma_{m,l}\rangle$ (with some $m,l$), and the overall shared state is also of a similar form. Thus, the swapped state can be transformed into the target state by a local unitary applied by Alice or Bob, removing the need for postselection.

To illustrate above statement in detail, we first focus on the case of a single intermediate node. Tracing the steps in Sec.~\ref{subsec:single}, the state before the swapping process is just $|\psi_T\rangle\otimes|\psi_T\rangle$. After the application of $U^\dagger$ and measurement in the node $1$, the resulting state for each separate measurement outcome $|\sigma_{m,l}\rangle$ becomes
\begin{align} \label{eq:mthswap2}
|\psi_{AB,m,l}^{(1)}\rangle = \frac{1}{\sqrt{p_{m,l}}} \sum_{i,j} \frac{1}{d_S}\, r_{ij,ml}\: |\lambda^A_i\rangle\otimes |\lambda^B_j\rangle,
\end{align}
where 
\begin{align}
    r_{ij,ml}&\equiv\langle \sigma_{m,l}|U^\dagger\left(|\lambda^A_j\rangle\otimes|\lambda^B_i\rangle\right)   \nonumber\\
    & = 
    \langle \psi_{m,l}|\left(|\lambda^A_j\rangle\otimes|\lambda^B_i\rangle\right) = \frac{e^{-2\pi ijl/d_S}}{\sqrt{d_S}}\delta_{i,j+m},\label{eq:coeff_proj}
\end{align}
and the probability of that outcome reads
\begin{equation}\label{eq:prob3}
p_{m,l}=\sum_{i,j}\frac{1}{d_S^2}|r_{ij,ml}|^2 
= \frac{1}{d_S^2}.
\end{equation} 
Since $\sum_{m,l=0}^{d_S-1}p_{m,l}=1$, we see that, indeed, the measurement ourcome is necessarily some of the states $|\sigma_{m,l}\rangle$ when using the the designed $U$. The swapped state corresponding to a measurement outcome $\sigma_{m,l}$ is
\begin{align} \label{eq:mthswap_final}
|\psi_{AB,m,l}^{(1)}\rangle = \frac{1}{\sqrt{d_S }} \sum_{j}  e^{-2\pi ijl/d_S} \: |\lambda^A_{j+m}\rangle\otimes |\lambda^B_{j}\rangle.
\end{align}
Each of these target-like states can be mapped into the target state \eqref{eq:final_m}, or equally $|\psi_{AB,0,0}^{(1)}\rangle$, by applying, for instance, local operator $U_{A,ml}$ by Alice who can be easily informed about measurement results at the intermediate node through classical communicates. The local unitary $U_{A,ml}$ which rotates Alice's basis  as $U_{A,ml}|\lambda^A_{j}\rangle=  e^{2\pi ijl/d_S}|\lambda^A_{j-m}\rangle$ will then map the shared state to the target $U_{A,ml}|\psi_{AB,m,l}^{(1)}\rangle=|\psi_T\rangle$. Analogously, Bob can carry out the local transformation instead of Alice.

It is straightforward to generalize the previous result to $m$ intermediate nodes. For the case $m=2$, the state prior to the second swapping is $|\psi_{AB}^{(1)}\rangle \otimes |\psi_T\rangle$. Following the same steps as outlined above, the state $|\psi_{AB}^{(2)}\rangle$ after swapping at the second node retains the same form as the single-node result in Eq.~\eqref{eq:mthswap_final}, with the indices $m$ and $l$ determined by the outcome of the second measurement. By iterating this procedure across $k$ intermediate nodes, the resulting state will always maintain the form of Eq.~\eqref{eq:mthswap_final}. In a straightforward approach, one would apply local unitaries $U_{A,ml}$ after each measurement at the intermediate nodes. However, this is not strictly necessary; it suffices to apply a single unitary at the final step. The reason is that, by repeating the above procedure without these post-measurement local unitaries, the resulting state takes the form
\begin{equation}
\label{eq:mthswap_k_node}
|\psi_{AB,M,L}^{(k)}\rangle = \frac{1}{\sqrt{d_S}} \sum_{j} e^{-2\pi i j L / d_S} \, |\lambda^A_{j+M}\rangle \otimes |\lambda^B_{j}\rangle,
\end{equation}
where $L = \sum_{i=1}^{k} l_i$ and
$M = \sum_{i=1}^{k} m_i$, with $l_i$ and $m_i$ corresponding to the measurement outcomes at node $i$. Therefore, a single final local unitary by Alice (or Bob) could map the shared into the target state with a local unitary $U_{A,ML}|\psi_{AB,M,L}^{(k)}\rangle=|\psi_T\rangle$.

\subsection{Swapping general states with unit fidelity without postselection} \label{subsec:general_nonpostselection}

In the above, we established that states with constant Schmidt coefficients can be shared across the network with unit fidelity without postselection. Building on these results, we now show that it is possible to share general states $|\psi_T\rangle$ of the form \eqref{eq:schmidt1} exactly, and again, without postselection. The key idea is to replace all the unitary operators $U$, except for the last one in the network, as well as all the $U^\dag$ operators in Fig.~\ref{fig:network}, with their corresponding unitaries that generate $d_S^2$ equal-weight superposition in their Schmidt decomposition of the form \eqref{eq:basis_shift_comp}
but with identical Schmidt eigenstates as in the target states.
We retain the generic unitary, denoted by $U$, whose action on $|\sigma_0\rangle$ produces a target state with a generally non-uniform Schmidt spectrum. The corresponding unitaries that yield equal-weight states $|\psi_{m,l}\rangle=\tilde{U}|\sigma_{m,l}\rangle$, as defined in Eq.~\eqref{eq:basis_shift_comp} are denoted by $\tilde{U}$.

To illustrate how this protocol enables perfect (unit-fidelity) state sharing without postselection, we first consider the simple case involving only Alice, Bob and a single intermediate node ($k=1$). 
First, Alice and the node apply $\tilde{U}$ and $U$, respectively, which leads to
\begin{align}
    |\Psi_1 \rangle &=
    \tilde{U} \otimes U(|\sigma_0\rangle\otimes|\sigma_0\rangle) \nonumber\\
    & = 
    \sum_{i,j=0}^{d_S-1} \sqrt{\frac{\lambda_j}{d_S}}\:|\lambda^A_i\rangle\otimes |\lambda^B_i\rangle\otimes|\lambda^A_j\rangle\otimes|\lambda^B_j\rangle.
\end{align}
Next, the operator $\tilde{U}^\dagger$ is applied at the node and followed by a measurement at the node, resulting in
\begin{align} \label{eq:mthswap2-nonuniform}
|\psi_{AB,m,l}^{(1)}\rangle = \frac{1}{\sqrt{p_{m,l}}} \sum_{i,j} \sqrt{\frac{\lambda_j}{d_S}}\, r_{ij,ml}\: |\lambda^A_i\rangle\otimes |\lambda^B_j\rangle.
\end{align}
It can be readily verified that $r_{ij,ml}$, for each measurement outcome, follows exactly the same form as in Eq.~\eqref{eq:coeff_proj}. Consequently, the corresponding probabilities are given by 
\begin{equation}\label{eq:prob3-nonuniform}
p_{m,l}=\sum_{i,j}\frac{\lambda_j}{d_S}|r_{ij,ml}|^2 
= \frac{1}{d_S^2}\sum_{j}\lambda_j = \frac{1}{d_S^2}.
\end{equation}
Substituting the expressions for the probabilities and coefficients $r_{ij,ml}$, the corresponding shared
states for each pair $(m,l)$ become
\begin{align} 
\label{eq:mthswap_final_nonuniform}
|\psi_{AB,m,l}^{(1)}\rangle =  \sum_{j}  \sqrt{\lambda_j }\:e^{-2\pi ijl/d_S} \: |\lambda^A_{j+m}\rangle\otimes |\lambda^B_{j}\rangle.
\end{align}
As in the case analyzed previously, these states differ from the exact target state 
$|\psi_T\rangle = \sum_{j}  \sqrt{\lambda_j } \: |\lambda^A_{j}\rangle\otimes |\lambda^B_{j}\rangle$
only by a local unitary operator applied by either Bob or Alice, depending on the measurement outcomes wich can be sent to either end by classical communication. Moreover, since the measurement outcome probabilities sum to one, the desired target state can be obtained exactly without postselection. Depending on each outcome, the swapped state can be mapped to the target state by applying an additional local operator, for instance, $U_{A,m,l}$ applied by Alice.

The generalization to the case involving additional intermediate nodes follows straightforwardly, in the same manner as discussed in the previous subsection. In Fig.~\ref{fig:network}, $\tilde{U}$ and $\tilde{U}^\dag$ are applied throughout the network, while only the final node $m$ employs the original unitary $U$, which produces the non-uniform Schmidt spectrum. Given a target state, determining both $U$ and $\tilde{U}$ is, in principle, always possible. However, for complex states, the task may become challenging. When the entanglement in the shared state is not high or only a couple of mediating nodes are needed, the many-body swapping protocol involving postselection, as formulated in the first part of this section, can be employed to share states with constant Schmidt coefficients with unit fidelity.

\section{Examples of many-body swapping}\label{Sec.3}

Here we illustrate the general many-body entanglement swapping protocol with examples. To set the stage, we provide a proof of concept of the protocol in real quantum hardware by sharing multi-qubit GHZ states. While this outcome could be accomplished by a standard Bell pair swapping and local operations, this example shows that the many-body protocol is robust to imperfections in the present quantum hardware. In Sec.~\ref{subsec:fault_tolerance} this example is also used to illustrate how the many-body protocol can readily implement quantum error correction and fault-tolerant swapping. The true versatility of the many-body swapping protocol, however, becomes evident when addressing general problems of sharing arbitrary many-body entangled states with non-uniform Schmidt spectra. In such cases, pair-swapping approaches become impractical, and the many-body protocol offers efficient alternative.

\subsection{Proof of concept on quantum hardware} \label{sec:GHZ}

\begin{figure}[t]
    \centering
    \includegraphics[width=.99\columnwidth]{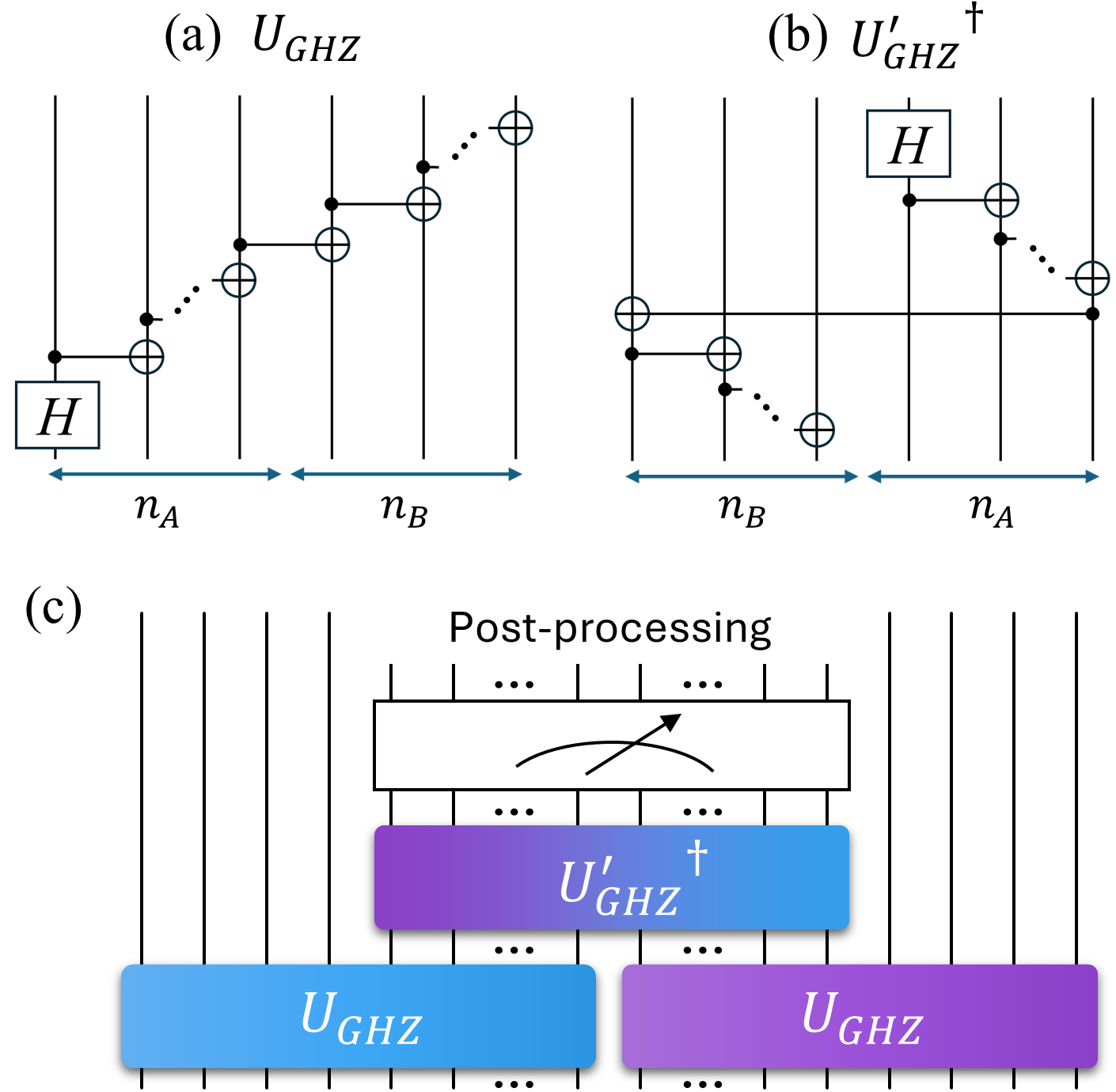}  
    \caption{  (a) Circuit for preparing GHZ states. (b) Inverse unitary in the basis where the order of Alice and Bob qubits are switched. (c) Protocol for creating a shared GHZ state for the outermost qubits.}
    \label{fig:swappingGHZ}
\end{figure}

Here we consider the task of distributing an $n$-qubit GHZ state between two parties, Alice and Bob. The target state is given by
\begin{align}\label{eq:ghz}
|\psi_{T}\rangle = \frac{1}{\sqrt{2}}(|00\ldots0\rangle + |11\ldots1\rangle),
\end{align}
where Alice and Bob initially each hold $n$ qubits and can perform arbitrary operations on their respective systems. To implement the protocol, both parties must locally prepare an $n$-qubit GHZ state. This is achievable from the initial state $|00\ldots0\rangle$ by a quantum circuit composed of a Hadamard gate followed by a series of CNOT gates, as shown in Fig.~\ref{fig:swappingGHZ}(a). The goal of the many-body entanglement swapping is to generate a shared GHZ state between $n_A$ qubits on Alice’s side and $n_B = n - n_A$ qubits on Bob’s side, while transferring the remaining qubits to an intermediary, Eve. Following the general procedure outlined in Fig.~\ref{fig1}, Eve applies a basis-reordered inverse unitary operation to the qubits received from Alice and Bob, illustrated in Fig.~\ref{fig:swappingGHZ}(b). The full circuit implementation of the protocol is shown in Fig.~\ref{fig:swappingGHZ}(c). Upon completing her operation, Eve measures her qubits. The possible outcomes are 
\[
\{|00\ldots0\rangle, |00\ldots,11\ldots\rangle, |11\ldots,00\ldots\rangle, |11\ldots1\rangle\},
\]
each occurring with probability $1/4$. If the result is $|00\ldots0\rangle$, the resulting state shared between Alice and Bob matches the target GHZ state $|\psi_{T}\rangle$, with Alice holding the first $n_A$ qubits. Interestingly, even the other measurement outcomes lead to GHZ-like entangled states, differing from $|\psi_{T}\rangle$ only by local bit-flip operations. This is indeed an example of a state with uniform Schmidt spectrum where postselection can be avoided, as the generating unitary is readily in the form that maps computational basis states to the states of form \eqref{eq:basis_shift_comp}. Thus, shared state can be transformed into the target state by local operations, in this case local Pauli gates $\sigma_x$ and $\sigma_y$.

\begin{figure*}[t]
    \centering
    \includegraphics[width=2\columnwidth]{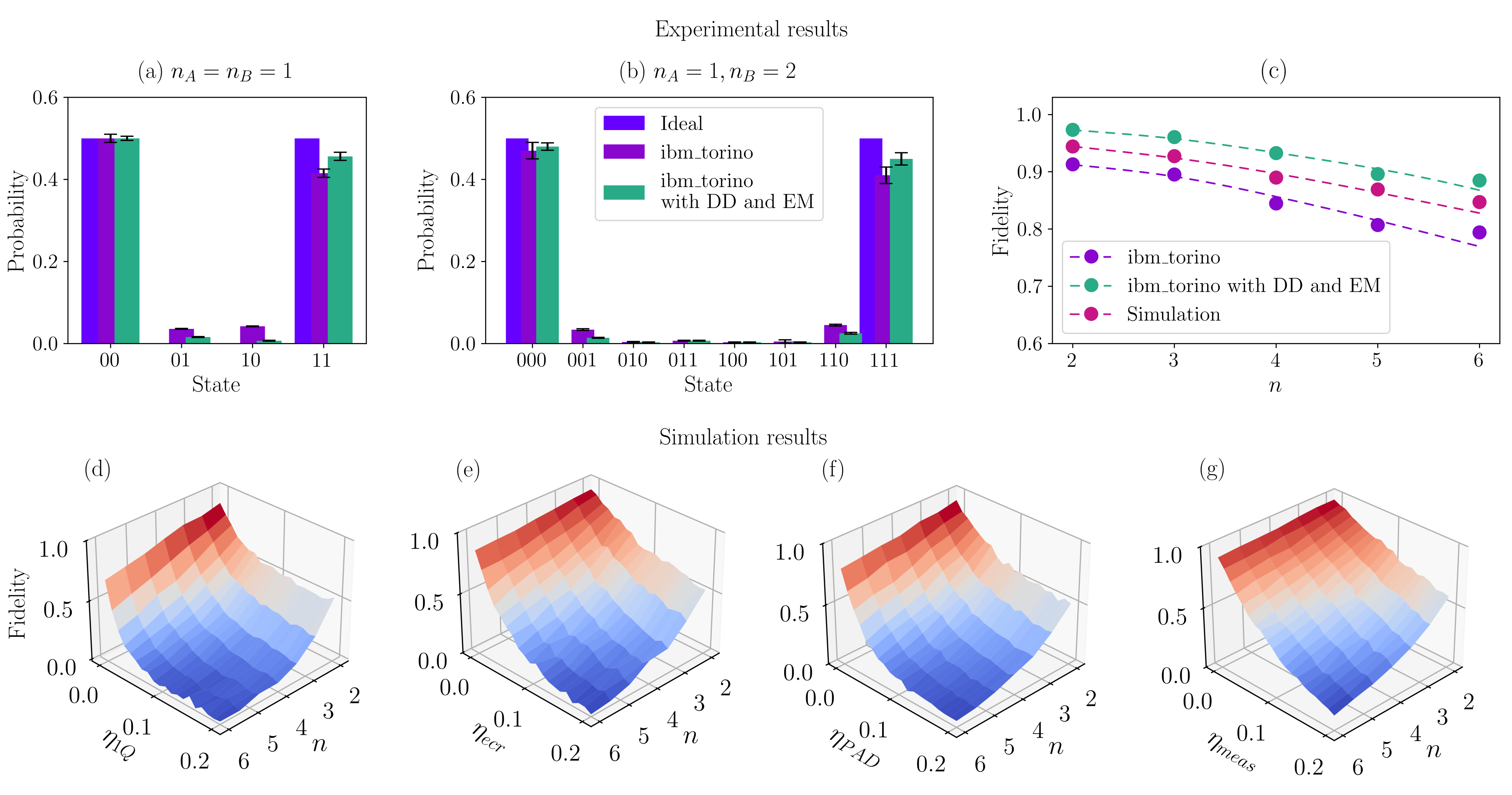}  
        \caption{(a–b) Probability distributions of vector states for the 2- and 3-qubit GHZ states obtained from experimental results on quantum hardware. The error bars represent the standard deviation. 
(c) GHZ state fidelity as a function of qubit number for three different approaches: (1) \textit{ibm\_torino}, (2) \textit{ibm\_torino} with dynamical decoupling (DD), error mitigation (EM), and (3) \textit{Simulation}. (d–g) Simulated fidelity decay of GHZ states as a function of qubit number under various noise sources: (d) single-qubit gate errors, (e) ECR gate errors, (f) phase-amplitude damping, and (g) readout errors.  
See Appendix~\ref{subsec:methods-1}, for detailed descriptions of the \texttt{ibm\_torino} hardware and the noise models used in the simulations.}
 
    \label{fig:ibm-result}
\end{figure*}

To demonstrate a proof of concept for our many-body entanglement-swapping protocol under realistic conditions, we implemented GHZ state sharing using IBM’s superconducting quantum hardware \cite{quantumprocessor}. The protocol was executed for systems of up to 12 qubits, enabling the sharing of GHZ states with up to 6 qubits (see App.~\ref{subsec:methods-1}, for details of the simulations on real hardware). The results are summarized in Fig.~\ref{fig:ibm-result}, which includes both experimental outcomes and corresponding simulations incorporating realistic noise models.
Figures~\ref{fig:ibm-result}(a) and (b) display the experimentally measured output distributions for Alice and Bob's qubits in GHZ states of $n = 2$ and $n = 3$ qubits, respectively. The observed bit-string frequencies exhibit the characteristic correlations of GHZ states, confirming the expected entanglement structure.  Figure \ref{fig:ibm-result}(c) presents the fidelity of the shared target state as a function of the number of qubits, comparing results obtained from the quantum hardware using three different approaches: 
(i) \textit{ibm\_torino}, corresponding to raw bitstring data collected from the ibm\_torino quantum processor, also without post-processing; and 
(ii) \textit{ibm\_torino} combined with additional post-processing, including dynamical decoupling and enhanced error mitigation strategies; (iii) \textit{Simulation}, which represents the raw bitstring outcomes obtained from the \emph{circuit simulator} \cite{javadi2024Qiskit}. It emulates the properties of real hardware but does not actually run on a quantum device, and includes no post-processing. 
 This comparison isolates the contribution of hardware-induced errors, demonstrating that standard error mitigation techniques can recover fidelity levels close to the performance ceiling imposed by the physical device.

Panels~(d-g) of Fig.~\ref{fig:ibm-result} present the simulated fidelity degradation as a function of qubit number under varying strengths of different noise sources, including single-qubit gate errors, two-qubit ECR gate errors, phase-amplitude damping, and readout errors (see Appendix~\ref{subsec:methods-2} for noise model details). These results identify the dominant noise mechanisms and provide valuable insight into the protocol’s scalability and robustness under realistic conditions. Overall, the data confirm that the many-body swapping protocol can be implemented with good fidelity on existing quantum hardware. Furthermore, as discussed in Sec.~\ref{subsec:fault_tolerance}, the protocol is compatible with standard quantum error correction techniques, allowing for a fault-tolerant implementation of entanglement swapping.

\begin{figure}[t]
    \centering \includegraphics[width=.99\columnwidth]{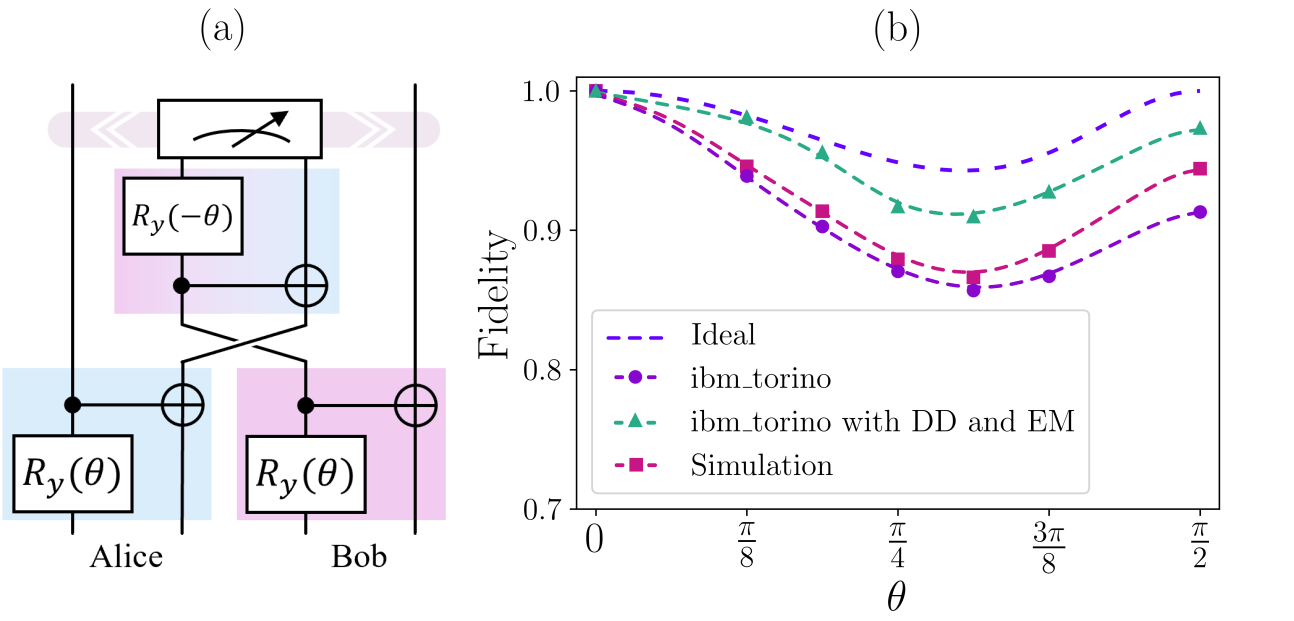}  
    \caption{Entanglement swapping of states $|\psi_T\rangle=\cos{\frac{\theta}{2}}|00\rangle+ \sin{\frac{\theta}{2}}|11\rangle$ which admits a non-uniform Schmidt spectrum for $0\leq\theta<\pi/2$. (a): Circuit for the swapping protocol. (b) Fidelity of the observed shared state as a function of angle $\theta$. The \textit{Ideal} line represent the theoretical prediction from Eq.~\eqref{eq:overlap2} } 
    \label{fig:nonuniform}
\end{figure}

Finally, to complement the simulation results for states with uniform Schmidt coefficients, Fig.~\ref{fig:nonuniform} illustrates the results for swapping target states of form  $|\psi_T\rangle=\cos{\frac{\theta}{2}}|00\rangle+ \sin{\frac{\theta}{2}}|11\rangle$ for different $\theta$. The Schmidt coefficients of the target states vary strongly as a function $\theta$, and the fidelity of the shared state can be compared to the theoretical prediction in Eq.~\eqref{eq:overlap2}.    The simulation results follow our theoretical prediction accurately, illustrating the protocol also for states with non-uniform Schmidt spectrum.

\subsection{Sharing entangled states with non-uniform Schmidt spectra}

We now consider sharing target states with non-uniform Schmidt spectrum. This class of states poses a challenge for pair-swapping methods and highlights the applicability of the many-body protocol. As a concrete example, we consider a four-qubit target state of the form $|\psi_T\rangle = U|0000\rangle$, generated by some unitary transformation $U$, where Alice and Bob each hold two qubits. The Schmidt rank for this bipartition is $d_S = 2^2 = 4$, and the state is generally characterized by four distinct Schmidt coefficients $\{\lambda_i\}$ with $i = 1, \dots, 4$.

To share such a state using the many-body protocol, we consider a representative case where the Schmidt coefficients are chosen as $\{\lambda_i\} = \{0.4, 0.3, 0.2, 0.1\}$. In this instance, the fidelity between the target state $|\psi_T\rangle$ and the shared state of Alice and Bob, denoted $|\psi_{AB}\rangle$, is found to be $F = \langle \psi_T | \psi_{AB} \rangle = 0.95$, as given by Eq.~\eqref{eq:overlap2}. The Schmidt coefficients of the resulting shared state, computed using Eq.~\eqref{eq:schmidt2}, are $\{\lambda_i^{AB}\} = \{0.64, 0.27, 0.08, 0.01\}$, indicating a distortion from the original spectrum. According to Eq.~\eqref{eq:prob1}, the probability of successful postselection in this example is $p_0 = 0.1$. This implies that, on average, the protocol must be repeated approximately ten times to successfully share a single copy of $| \psi_{AB} \rangle$. While the shared state does not match the target exactly, the high fidelity underscores the protocol’s effectiveness. A high fidelity can persist even for much larger and more complex states, as seen in the following example.

\subsection{Sharing multi-qubit states generated by random unitary circuits}
\begin{figure}[t]
    \centering \includegraphics[width=.99\columnwidth]{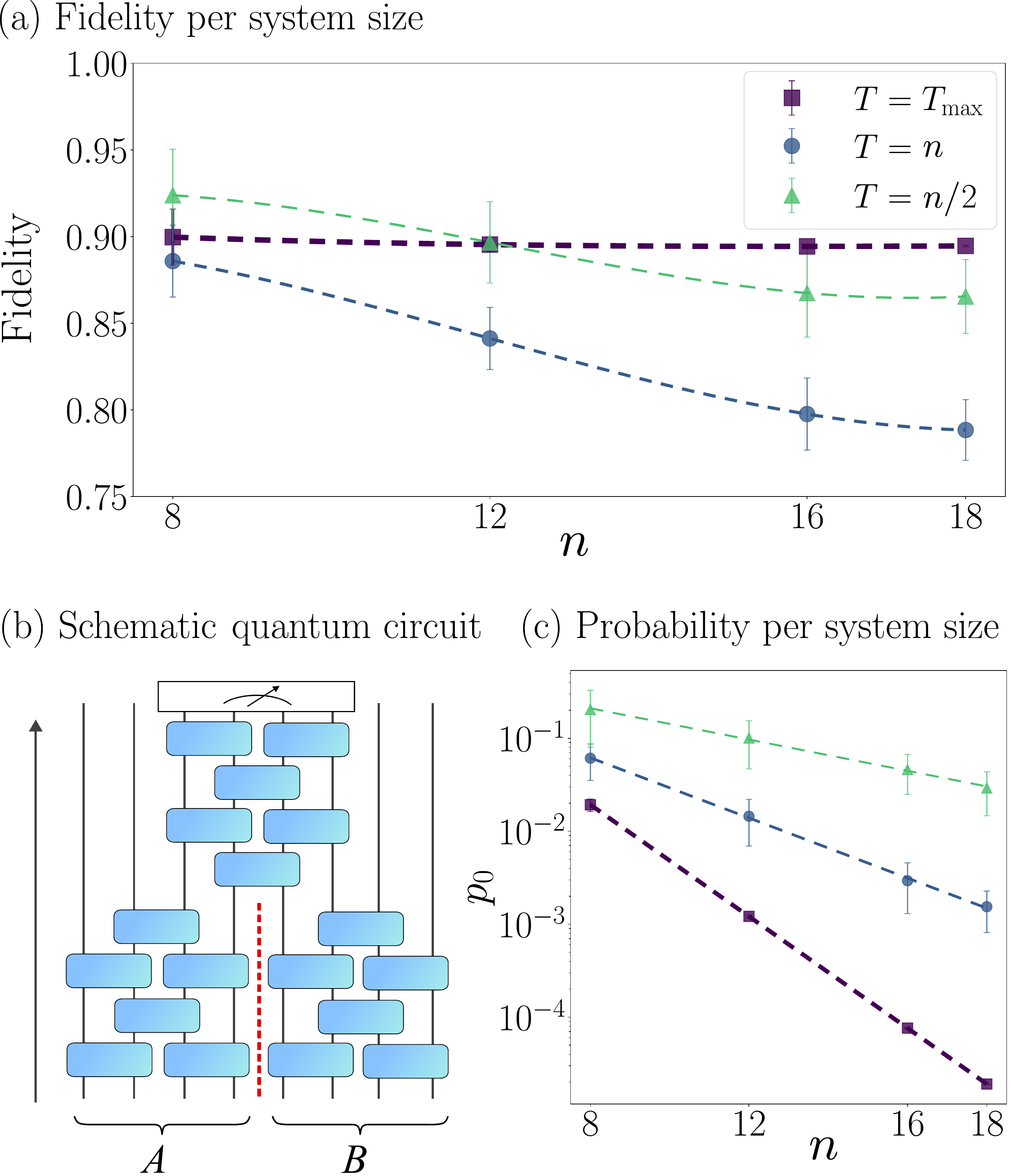}  
    \caption{Sharing states generated by a random unitary circuit:
    (a) The average fidelity (overlap of the shared state with the target state) 
    given by expression \eqref{eq:overlap2} 
    as a 
    function of the system size $n$ for random two-qubit brick circuits. 
    (b) Schematic illustration of the random unitary circuit generating the target state shared between Alice and Bob.
    (c) The postselection probability of obtaining the desired shared state as a function of the system size. 
    }
    \label{fig:random_unitary}
\end{figure}

Next we illustrate how the many-body swapping protocol enables high-fidelity sharing of complex, high-entropy quantum states. In particular, we focus on target states generated by random two-qubit brickwork circuits, as depicted in Fig.~\ref{fig:random_unitary}(b). Such circuits have become widely used as testbeds for exploring a range of many-body quantum phenomena, especially in the study of entanglement dynamics and information scrambling \cite{Nahum2017,Fisher2023review,Minnich_2023measurement}. In the model considered here, the time-evolution operator for each cycle is defined as
\begin{equation} 
\mathbf{u}(t) = \prod_{\text{even } l} u_{l,l+1}(2t) \prod_{\text{odd } l} u_{l,l+1}(2t - 1),\nonumber
\end{equation}
where $u_{l,l+1}(\tau)$ denotes a two-qubit unitary acting on neighboring qubits $l$ and $l+1$ at the discrete time step $\tau = 2t$ or $2t - 1$. Each two-qubit gate is generated as
\begin{equation} 
u_{l,l+1}(t) = e^{ i \left(\phi_x X_l X_{l+1} + \phi_y Y_l Y_{l+1} + \phi_z Z_l Z_{l+1}\right)},\nonumber
\end{equation}
where $X_l$, $Y_l$, and $Z_l$ are the Pauli operators on qubit $l$, and the coefficients $\phi_i$ are independent random variables uniformly sampled from the interval $[-\pi, \pi]$. These random parameters are independently drawn for each gate location and time step. This ensemble of unitaries does not sample from the full Haar measure over two-qubit unitaries. However, the brick circuits generate maximum entropy states, up to finite-size corrections. Notably, when acting on an initial product state, these circuits rapidly generate entanglement across subsystems. After order of $T \sim n_s$ layers, where $n_s$ is the number of qubits in the shared state, the entanglement entropy saturates close to the maximal value, scaling as $S_{n_s}/(n_s \ln 2) \sim 1 + \mathcal{O}(1/n_s)$. 
The fact that the random brickwork circuits generate nearly maximally entangled states has attracted significant interest in the past decade. Below we see that the highly scrambled, complex quantum states produced by such circuits can be shared by the many-body swapping protocol with high fidelity.

As seen in Fig.~\ref{fig:random_unitary}, for deep circuits ($T=T_{\rm max} \gtrsim 2n$) the average fidelity of the shared state, given by  Eq.~\eqref{eq:overlap2}, is strikingly high.  As the total system size increases from $n = 4$ to $n = 18$, the fidelity stays around $F = 0.9$. This is quite remarkable, given the complexity of these nearly maximum entropy states. However, as seen in Fig.~\ref{fig:random_unitary} (c), the state sharing is exponentially costly in the system size, as the success probability for postselection is determined by the 3rd R\'enyi entropy through Eq.~\eqref{eq:prob1}. For highly entangled states approaching maximal entropy, the postselection probability scales approximately as $p_0 \approx 2^{-n}$. For instance, successfully sharing of a single $n = 18$ state (split evenly between two parties) would require on average $2.6\times10^5$ repetitions. This example highlights the substantial resource overhead associated with distributing highly entangled, high-entropy quantum states.  Still, sharing smaller states $n\leq 10$ would require less than $10^3$ repetitions on average.

\subsection{Application of many-body swapping in distributed quantum computing} \label{subsec:DistributedQC}

\begin{figure}[t]
    \centering \includegraphics[width=.99\columnwidth]{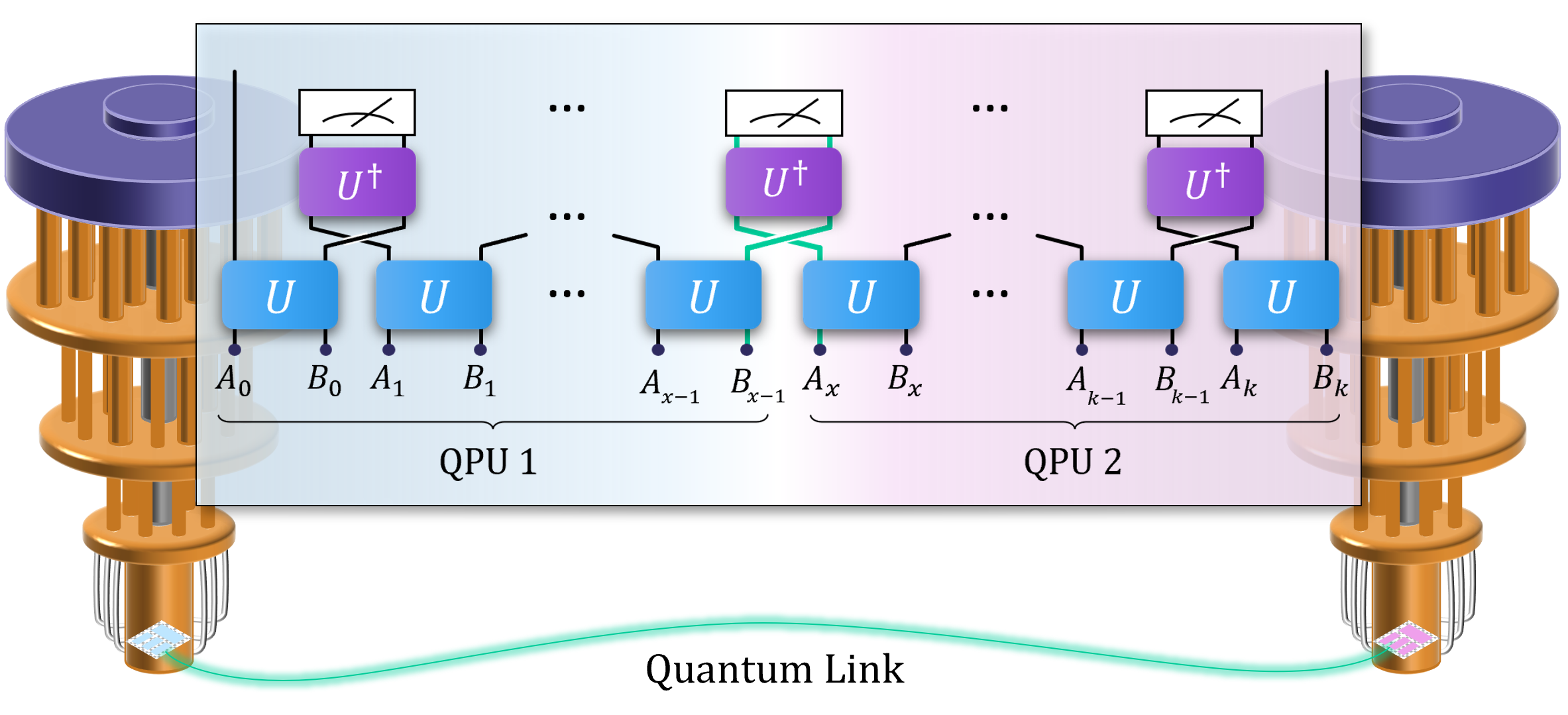}  
    \caption{Quantum state sharing between different quantum processors nodes and processors. Some nodes in the network can physically reside on the same chip, some can lie in different cryogenic environments coupled by a quantum link.}  
    \label{fig:DQC}
\end{figure}

In this section, we elucidate how the swapping protocol, with multiple intermediate nodes, can be employed in distributed quantum computing. Here Alice and Bob represents different quantum processor nodes or separate modular subunits within larger quantum computing architecture, as illustrated in Fig.~\ref{fig:DQC}. The far-apart processor nodes are connected via chain of other nodes, each coupled to their nearest neighbors. This array of processor nodes can be regarded as a linear network, and the goal here is to establish efficient quantum-state sharing between processor nodes which are not nearest neighbors. For this application, it is natural to adopt a slightly modified notation. Similar to the approach introduced at the beginning of Sec.~\ref{subsec:single}, each node, including those of Alice and Bob, is divided into two parts labeled $A_i$ and $B_i$, which together constitute node $i$ in Fig.~\ref{fig:network}. Accordingly, $i=0$ corresponds to Alice’s node and $i=k$ to Bob’s node, resulting in a total of $2(k+1)$ distinct parts.

In this framework, the goal is to share a state between parts $A_0$ and $B_k$. A set of unitary operators $U$ is applied to the corresponding pairs $(A_i, B_i)$ for $i \in (0, \cdots, k)$. Subsequently, the operators ${U^\prime}^\dag$ are applied to the pairs $(B_i, A_{i+1})$ for $i \in (0, \cdots, k-1)$, each followed by measurement of all qubits in the parts $B_i$ and $A_{i+1}$. As a result, the state becomes distributed between spatially separated parts without any direct quantum link (unitary gate) connecting them. This is accomplished by applying semi-local operators only between neighboring parts and by measuring all intermediate parts. Formally, the state distribution achieved through successive semi-local operations can be expressed as
\begin{equation}\label{eq:state-teleportation}
    |\psi_{A_0,B_k}\rangle = 
    \prod_{i=1}^{k} 
    {\mathbb M}^{\bm 0}_{B_{i-1},A_{i}}
    {U_{A_{i},B_{i-1}}^{\dag}}
    \:
    U_{A_{i},B_{i}} \: U_{A_0,B_0}
    |\bm{0}\rangle_{\rm all},
\end{equation}
where $|\bm{0}\rangle_{\rm all}$ denotes the initial state in which all qubits across all nodes are prepared in the $0$ state. The indices of the unitary operators indicate the subsystems over which these multi-qubit gates are applied. The measurement operators are denoted by ${\mathbb M}^{\bm 0}_{(B_{i-1},A_{i})}\equiv |\bm{0}\rangle\langle\bm{0}|/\sqrt{p_{\bm 0}}$, where $|\bm{0}\rangle$ corresponds to the all-zero product state of the combined subsystems ${B_{i-1},A_{i}}$, and $p_{\bm 0}$ represents the probability of obtaining that measurement outcome. 

The above version assumes carrying out postselection of measurement outcomes which, as discussed in Secs.~\ref{subsec:postselection}, \ref{subsec:general_nonpostselection}, can be avoided by a tailored feedforward method. In that case, the distributed state can be rewritten in a postselection-free form as
\begin{align}\label{eq:state-teleportation-without-postselection}
    |\psi_{A_0,B_k}\rangle = U^{\{{\bm m}\}}_{A_0,{\rm loc}}
    &\:\prod_{i=1}^{k} 
    {\mathbb M}^{{\bm m}_i}_{B_{i-1},A_{i}}
    \nonumber\\
    &
    {U_{A_{i},B_{i-1}}^{\dag}}
    U_{A_{i},B_{i}} \: U_{A_0,B_0}
    |\bm{0}\rangle_{\rm all},
\end{align}
where ${\bm m}_i$ labels the measurement outcomes obtained from measuring the qubits in the parts $B_{i-1}$ and $A_{i}$. As explained in Sec.~\ref{subsec:postselection}, the final local unitary operator $U^{\{{\bm m}\}}_{A_0,{\rm loc}}$ restores the system to the desired target state, depending on the complete sequence of measurement outcomes $\{{\bm m}\} = ({\bm m}_1,{\bm m}_2,\cdots,{\bm m}_k)$. In Secs.~\ref{subsec:postselection} and \ref{subsec:general_nonpostselection}, the measurement outcomes were labeled with two indices, so ${\bm m}$ here corresponds to the combined indices $(m,l)$ used previously.

Equations \eqref{eq:state-teleportation} and \eqref{eq:state-teleportation-without-postselection} illustrate that in this approach, there might not be a need for a physical transfer of quantum data in the sense of long-range quantum communication. It suffices to be able to perform semi-local quantum operations on two neighboring parts. Considering, for example, superconducting quantum hardware, these $2k+2$ parts may reside in different quantum processing units (QPUs) on the same chip or different cryogenic environments, provided that the operators $U$ or ${U^\prime}^\dag$ can be applied between neighboring parts, where one part belongs to one QPU and the other to another. Such inter-chip quantum links are not only feasible in principle but have already been demonstrated between physically separate superconducting qubits connected via cryogenic waveguides~\cite{Wallraff2020,caleffi2024distributed}. Conceptually, some of these parts can be regarded as distinct QPUs connected through quantum links that enable inter-chip unitary operations. For instance, in a full network comprising two QPUs, only a single inter-chip quantum link, as shown in red in Fig.~\ref{fig:DQC}, is required to implement the boundary unitary operation.
Therefore, the scheme provided here can be used in efficiently distributing complex states and quantum computating over multiple processing units which are in a sense locally linked to each other.

\section{Discussion}

Here we highlight two crucial advantages of the many-body swapping protocol in comparison to the existing entanglement swapping methods.  


\subsection{Quantum state sharing with the many-body protocol versus pair swapping}

Here we make the case why the many-body swapping protocol provides more practical approach for high-fidelity sharing of general target states between non-signaling parties compared to the existing pair-based swapping methods. In the pair-based approaches, Alice and Bob aim to share a general $n$-qubit target state, where each party ends up holding $n/2$ qubits. They begin with a number of shared (maximally or partially) entangled pairs and the ability to perform local operations.

To understand the practical limitations of the pair-swapping methods, we note that upon bipartitioning a generic target state between Alice and Bob, the number of different Schmidt values is bounded by the Schmidt rank as $d_S - 1$, hence, can be as large as $2^{n/2} - 1$. Since local unitary operations preserve the Schmidt coefficients, and local measurements cannot independently adjust them, achieving a target state with $N_{\rm ind}$ distinct Schmidt coefficients would require access to at least $N_{\rm ind}$ entangled pairs— each carefully prepared with a unique, precisely tuned Schmidt coefficient. However, as pointed out above, for a generic many-qubit state, $N_{\rm ind}$ grows exponentially with the number of qubits and reaches up to $2^{n/2} - 1$. This implies that an exponentially large number of shared pair states would be needed, each tailored to contribute to a different component of the target state's entanglement structure. Moreover, even if such a resource were available, one would still face the profound problem of determining the local operations which map the collection of entangled pairs to the target state. In general, these local operations are a priori unknown and state specific. 

Due to these challenges, there is currently no known protocol based on pair-swapping that can reliably generate even a few-body target states with a generic Schmidt spectrum. In contrast, the many-body entanglement swapping protocol introduced in this work is explicit, straightforward to implement, and leads directly to a shared state with the correct Schmidt vectors and typically high fidelity.
Thus, given the formidable practical challenges associated with pair-swapping approaches, the many-body protocol stands out as an attractive and feasible method for sharing general entangled states between non-signaling parties.

\subsection{Error correction and fault-tolerant entanglement swapping } \label{subsec:fault_tolerance}

The simple GHZ state sharing example in Sec.~\ref{sec:GHZ} suggests an interesting application of the many-body swapping. While a shared GHZ could be generated by a single swapped Bell pair and local operations, the many-body protocol has an important advantage. Let's consider, for example, a realistic scenario in which Eve's measurement process is unideal. In that case, she has a small but finite probability $p^*$ of registering a wrong outcome from a measurement process of a single qubit. In pair swapping-based schemes, a single error in the swapping measurement propagates to the shared GHZ state and contaminates it with the same probability $p^*$. On the other hand, in the many-body swapping protocol, Alice and Bob send multiple qubits to Eve and a single error in the measurement process can be detected. For instance, instead of obtaining a correct outcome $|0000\ldots1111\rangle$, Eve could in fact register an erroneous outcome  $|0000\ldots1101\rangle$, where a single bit has flipped. Despite the error, Eve has no difficulties in identifying the state correctly and communicating the right result to Alice and Bob.

The error detection example above reflects the fact that a GHZ state can be regarded as a Bell state formed by logical qubits which are encoded by a simple repetition code $|0\rangle \to |00\ldots\rangle$ and $|1\rangle \to |11\ldots\rangle$. This is a primitive example of how a single logical qubit can be encoded to multiple physical qubits using quantum error-correcting codes (QECCs). As mentioned in the general derivation and illustrated by the GHZ example above, the many-body entanglement swapping protocol is agnostic to the basic building blocks of the target state, applying equally well for physical qubits as well as blocks of physical qubits encoding a logical qubit. This means that the standard machinery of quantum error correction can be readily accommodated in the protocol, opening the door to fault-tolerant entanglement swapping. 

Logical qubits can be encoded using established quantum error-correcting codes (QECCs), such as stabilizer codes, allowing our protocol to share any target state preparable via stabilizer circuits \cite{terhal2015,acharya2024google}. This includes key classes of entangled states such as logical Bell pairs and logical cluster states. To extend the protocol toward fully fault-tolerant entanglement swapping for arbitrary target states, including those requiring non-Clifford operations, QECCs can be combined with magic state distillation techniques \cite{Litinski2019magicstate}. In this broader setting, if the generating unitary $U$ and Eve's projective measurements are implemented fault-tolerantly, the many-body entanglement swapping protocol can be executed in a fully fault-tolerant manner.

Compared to single-pair swapping, a fault-tolerant realization requires more advanced error-correction methods and efficient handling of multiple flying qubits, topics that have been explored in previous works \cite{nickerson2013topological,muralidharan2016optimal}. Nevertheless, in distributed quantum computing, many-body swapping offers important advantages over standard approaches that rely on numerous single-pair swappings combined with entanglement purification to generate a single high-quality Bell pair. Since large-scale processors already employ standard error-correction routines with syndrome measurements and correction operations, it is natural to assume that such capabilities are available in the processor modules and intermediate nodes performing many-body swapping (see Fig.~7).

Even for sharing logical Bell pairs between modules, many-body swapping can remove the need for repeated pair generation, entanglement purification at each node, and post-generation logical encoding. Furthermore, in distributed quantum computing architectures, restricting communication to single-pair sharing and single-qubit teleportation may not always be optimal, as directly sharing more complex entangled states can reduce overall computation time and enhance architectural flexibility. When quantum gates between nodes remains within the physical error threshold of the chosen QECC, many-body swapping becomes feasible and represents a significant step toward fully error-resilient distributed quantum systems.

\section{Summary and outlook}

We established a general many-body entanglement swapping protocol that enables two non-signaling parties to share a high-fidelity copy of a generic many-body quantum state, and we demonstrated its proof of principle on actual quantum hardware. In the emerging era of distributed quantum information processing and quantum networks, this protocol provides key functionalities that are difficult or impossible to achieve by other known methods. These include the sharing of complex states with a generic Schmidt spectrum and the implementation of fault-tolerant entanglement swapping. In forthcoming work, we will employ existing quantum hardware to demonstrate further aspects of the protocol, including sharing more complex states  \footnote{Y. Mafi, A. G. Moghaddam and T. Ojanen, in preparation}.

\section{Acknowledgements} 
S.H. and Y.M. contributed equally to this work. A.G.M. and T.O. acknowledge Jane and Aatos Erkko Foundation for financial support. T.O. also acknowledges the Finnish Research Council project 362573.

\appendix

\section{Simulation details on the quantum processor}
\label{subsec:methods-1}

To carry out the quantum simulations presented in Fig.~\ref{fig:ibm-result}, we used IBM’s 133-qubit superconducting quantum processor \texttt{ibm\_torino}, accessed via the IBM Quantum cloud platform \cite{quantumprocessor}. This device is based on fixed-frequency transmon qubits arranged in a heavy-hex lattice architecture. The processor exhibits coherence times ($T_1$ and $T_2$) typically ranging from 100 to 200~$\mu$s, with single- and two-qubit gate fidelities exceeding 99.9\% and 99\%, respectively, and and median readout error of $\sim3.21\%$. For benchmarking and validation purposes (data shown in Fig. \ref{fig:ibm-result}~(c)), we employed IBM’s high-performance simulators available through Qiskit \cite{javadi2024quantum}. These simulators realistically emulate IBM Quantum devices by incorporating noise models that account for native gate errors, qubit connectivity, measurement errors, and decoherence.

 We utilized the \emph{Sampler} measurement interface \cite{javadi2024quantum}. The Sampler executes quantum circuits repeatedly to produce bitstring distributions, faithfully capturing raw hardware noise without any post-processing. To mitigate the effects of noise and decoherence, we applied two readily available techniques: \emph{Matrix-free Measurement Mitigation} (M3) \cite{M3ref} and \emph{Dynamical Decoupling} (DD) \cite{viola1998dynamical, viola1999dynamical, duan1999suppressing}. M3 operates within a reduced subspace defined by the noisy input bitstrings, 
making the associated linear system much easier to solve because the number of unique bitstrings is far smaller than the full Hilbert-space dimension. DD, on the other hand, helps preserve quantum coherence during idle periods by applying carefully designed pulse sequences that dynamically refocus the qubit states.

The number of measurement shots, denoted by $m$, was selected to resolve probability differences down to a standard deviation of approximately $\sigma \sim 1\%$. The standard deviation in estimating a probability $p$ is approximated by $\sigma \approx \sqrt{p(1 - p)/m}$. 
To quantify the similarity between the output and target distributions, as the fidelities shown in Fig. \ref{fig:ibm-result}(d-g), we have adapted Hellinger distance $H$. Based on this, the fidelity is defined as $F = (1 - H^2)^2$ which measure lies in the interval $[0, 1]$ and is equivalent to the \emph{classical fidelity} $F_H$, since it coincides with the \emph{quantum state fidelity} for diagonal density matrices. Such classical fidelity between the output state $|\psi_{AB}\rangle$ and the target state $|\psi_T\rangle$ is given by:
\begin{align}
F_H(|\psi_{AB}\rangle, |\psi_T\rangle) = \left( \sum_i \sqrt{p_i q_i} \right)^2,    
\end{align}
where $p_i$ and $q_i$ are the probability components of the output and target distributions, respectively.

\section{Quantum noise models}\label{subsec:methods-2}

To model the noise effects, we adopt the Kraus representation of quantum channels. In this framework, a noisy quantum process acting on a density matrix $\rho$ is described by a completely positive, trace-preserving map $\mathcal{E}$, which can be expressed as:
\begin{equation}
    \mathcal{E}(\rho) = \sum_k E_k \rho E_k^\dagger,
\end{equation}
where the operators $\{E_k\}$ are known as Kraus operators and satisfy the completeness relation $\sum_k E_k^\dagger E_k = I$. In the following, we provide the explicit forms of the Kraus operators corresponding to different types of quantum errors, which are commonly encountered in quantum devices.

\emph{Single-qubit gate error}:
For single-qubit gate error, we are using a general Pauli channel error (depolarizing channel error) representation that includes $X$, $Y$, and Z error simultaneously, which applies Pauli errors $X$, $Y$, and $Z$ with certain probabilities:
\begin{equation}
    \mathcal{E}_{1Q}(\rho)=p_xX\rho X+p_yY\rho Y+p_zZ\rho Z+(1-\sum_{i\in\{x,y,z\}}p_i)\rho
\end{equation}
where $p_x,p_y, p_z\in[0,1]$ denote probability of applying $X$, $Y$, and $Z$ error, respectively. Hence, The Kraus operators for this channel are:
\begin{equation}\label{eqerror1Q}
    \begin{split}
        E_0=\sqrt{p_x}X, &\ \ \ E_1=\sqrt{p_y}Y,\\
        E_2=\sqrt{p_z}Z, &\ \ \ E_3=\sqrt{(1-\sum_{i\in\{x,y,z\}}p_i)}I.
    \end{split}
\end{equation}

The specific probability of $X$, $Y$, and $Z$ errors on single-qubit gates in IBM Quantum devices are not publicly detailed in terms of individual Pauli error components. IBM typically reports overall single-qubit gate error rates, which encompass various error sources, including decoherence and control imperfections. For simulation purposes, it's common to model single-qubit errors using a depolarizing channel, where $X$, $Y$, and $Z$ errors are assumed to occur with equal probability. Hence, we can consider the single-qubit gate error as 
\begin{equation}
\begin{split}
    \mathcal{E}_{1Q}(\rho)=&\frac{\eta_{1Q}}{4}X\rho X+\frac{\eta_{1Q}}{4}Y\rho Y\\
    & +\frac{\eta_{1Q}}{4}Z\rho Z+(1-\frac{3\eta_{1Q}}{4})\rho
\end{split}
\end{equation}
where $p_x=p_y=p_z=\frac{1}{4}\eta_{1Q}$.

\emph{Readout error}: For each qubits, the readout error can be implemented using a confusion matrix by applying after quantum circuit execution classically:
\begin{equation}
R=
    \begin{bmatrix}
        P(0|0)&P(0|1)\\
        P(1|0)&P(1|1)\\
    \end{bmatrix}=
    \begin{bmatrix}
        1-p_{01}&p_{01}\\
        p_{10}&1-p_{10}\\
    \end{bmatrix}
\end{equation}
where $P(i|j)$ is the probability to read $i$ given true outcome $j$. Therefore, $p_{01}$ and $p_{10}$ represent the probability of flipping $0\rightarrow1$ and $1\rightarrow0$, respectively.

\emph{Phase- and Amplitude-damping error}:
The pure dephasing time $T_\phi$ satisfy the following expression 
\begin{equation}
    \frac{1}{T_2}=\frac{1}{2T_1}+\frac{1}{T_\phi}
\end{equation}
where $T_1$ and $T_2$ are energy relaxation time and dephasing time, respectively.

Let $\gamma=1-e^{-t/T_1}$ be amplitude-damping channel probability, where $t$ is the gate duration time, and Kraus operators are
\begin{equation}
    \begin{split}
        E^{AD}_0 = \begin{bmatrix}
            1&0\\
            0&\sqrt{1-\gamma}
        \end{bmatrix},&\ \ \
        E^{AD}_1 = \begin{bmatrix}
            0&\sqrt{\gamma}\\
            0&0
        \end{bmatrix}
    \end{split}
\end{equation}
where $\gamma$ represents the probability that the qubit decays from $|1\rangle$ to $|0\rangle$.

The phase-damping error describes a noise process that is uniquely quantum mechanical, depicting the loss of quantum information without loss of energy. By considering $\lambda=1-e^{-t/T_{\phi}}$, the Kraus operators are
\begin{equation}
    \begin{split}
        E^{PD}_0 = \sqrt{1-\lambda}\cdot I,&\ \ \
        E^{PD}_1 = \sqrt{\lambda}\cdot Z
    \end{split}
\end{equation}
where $\lambda$ can be interpreted as the probability that a qubit loses quantum information without loss of energy.

The phase- and amplitude-damping error (PAD) can be consider as product of each noises, $\mathcal{E}_{PAD}(\rho) = \mathcal{E}_{PD}(\rho)\cdot \mathcal{E}_{AD}(\rho)$.
Which Kraus operators are
\begin{equation}
    \begin{split}
        E^{PAD}_0=E^{PD}_0\cdot E^{AD}_0\\
        E^{PAD}_1=E^{PD}_1\cdot E^{AD}_1
    \end{split}
\end{equation}
where $\eta_{PAD}$ denotes the probability of a qubit capturing both energy relaxation (amplitude damping) and loss of quantum coherence (phase damping).

\emph{ECR error}:
For the ECR gate error, similar to the single-qubit gate, we use a general Pauli channel error model, which can be expressed as a tensor product of single-qubit error representations \eqref{eqerror1Q},
\begin{equation}
    \mathcal{E}_{ecr}(\rho) = \sum_{k}{E^{ecr}_k\rho {E^{ect}_k}^\dagger}=\sum_{i,j}{(E_i\otimes E_j)\rho(E_i^\dagger\otimes E_j^\dagger)},
\end{equation}
and $\eta_{ecr}$ represents the effective error probability for the ECR gate, assuming the error contributions from the two input qubits are equal.

\bibliography{bibliography}
\end{document}